# Orbital Stability Close to Asteroid 624 Hektor using the Polyhedral Model

Yu Jiang[1, 2], Hexi Baoyin[1], Hengnian Li[2]

1. School of Aerospace Engineering, Tsinghua University, Beijing 100084, China
2. State Key Laboratory of Astronautic Dynamics, Xi'an Satellite Control Center, Xi'an 710043, China

Y. Jiang (✉) e-mail: jiangyu_xian_china@163.com

H. Baoyin (✉) e-mail: baoyin@tsinghua.edu.cn.

**Abstract**. We investigate the orbital stability close to the unique L4-point Jupiter binary Trojan asteroid 624 Hektor. The gravitational potential of 624 Hektor is calculated using the polyhedron model with observational data of 2038 faces and 1021 vertexes. Previous studies have presented three different density values for 624 Hektor. The equilibrium points in the gravitational potential of 624 Hektor with different density values have been studied in detail. There are five equilibrium points in the gravitational potential of 624 Hektor no matter the density value. The positions, Jacobian, eigenvalues, topological cases, stability, as well as the Hessian matrix of the equilibrium points are investigated. For the three different density values the number, topological cases, and the stability of the equilibrium points with different density values are the same. However, the positions of the equilibrium points vary with the density value of the asteroid 624 Hektor. The outer equilibrium points move away from the asteroid's mass center when the density increases, and the inner equilibrium point moves close to the asteroid's mass center when the density increases. There exist unstable periodic orbits near the surface of 624 Hektor. We calculated an orbit near the primary's equatorial plane of this binary Trojan asteroid; the results indicate that the orbit remains stable after 28.8375 d.

## 1 Introductions

To date there are only a few Jupiter binary Trojan asteroids. These are (617) Patroclus (Mueller et al. 2010; Nesvorny et al. 2010), (624) Hektor (Kaasalainen et al. 2002), (17365) 1978 VF11 (Mann et al. 2007; Noll et al. 2014), and (29314) Eurydamas (Mann et al. 2007; Noll et al. 2014), etc. Among these Jupiter binary Trojan asteroids (617) Patroclus, (17365) 1978 VF11, and (29314) Eurydamas are located on the L5-point of the Sun-Jupiter system. Only the binary asteroid system (624) Hektor is located on the L4-point of the Sun-Jupiter system. Among these binary Trojan asteroids, (624) Hektor is the largest one; the primary's diameter is 184 km (Marchis et al. 2014). (624) Hektor is the binary asteroid system which has elongated/bi-lobed primaries with a small moonlet and a rapidly rotating primary (Noll et al. 2014). In addition observational data for the irregular shape of asteroid (624) Hektor has been obtained in the literature Kaasalainen et al. (2002).

Prockter et al. (2005) investigated the PARIS (Planetary Access with Radioisotope Ion-drive System) spacecraft with the goal to explore the largest Jovian Trojan binary asteroid (624) Hektor. The candidate payload carried on this spacecraft includes cameras, a UV-Vis-IR spectrograph, etc. (Gold et al. 2007).

Due to these characteristics the L4-point Jupiter binary Trojan asteroid (624) Hektor has already received particular attention. Cruikshank et al. (2001) studied the composite spectrum of (624) Hektor, and suggested the presence of silicate minerals or the macromolecular carbon-rich organic material. Marchis et al. (2006) reported on the images of (624) Hektor which indicated the presence of a moonlet S/2006(624)1, and they estimated the diameter of the moonlet to be 15 km. Hamanowa and Hamanowa (2009) reported the synodic rotation period of (624) Hektor to be $6.9210 \pm 0.0001$ h. Marchis et al. (2012) estimated the semi-major axis of the moonlet between 1,100 and 1,400 km and the eccentricity between 0.13 and 0.20. The density of the primary was estimated between 1.8 and 3.5 g·cm$^3$ (Marchis et al. 2012). Marchis et al. (2014) investigated the puzzling mutual orbit of 624 Hektor, and suggested that the binary system of 624 Hektor and its moonlet Skamandrios has not evolved significantly due to tidal interactions. Doressoundiram et al. (2016) suggest a relatively homogenous surface for 624 Hektor; however, the possibility of localized inhomogeneity cannot be excluded. Rozehnal et al. (2016) used a smoothed particle hydrodynamics method and demonstrated that the moonlet can be created by a single impact event. Sharma (2016) applied volume averaging to analyze the stability of (624) Hektor. The results revealed that the most probable scenario for the primary is a contact binary which is joined by a force-carrying 'bridge', and a separated primary as a stable binary cannot be found.

However, the above literature mainly focuses on the physical parameters of the unique asteroid (624) Hektor. For other binary asteroid systems and triple asteroid systems, such as 243 Ida and 216 Kleopatra, there have been several investigations of the gravitational potential, equilibrium, and orbital stability (Wang et al. 2014; Chanut et al. 2015; Jiang et al. 2015). The dynamical environment for this binary Trojan asteroid is unknown. Thus in this paper we want to investigate the orbital stability close to asteroid 624 Hektor by considering its irregular shape. The shape data of asteroid 624 Hektor (Kaasalainen et al. 2002) include 2038 faces and 1021 vertexes. Figure 1 shows the 3D convex shape of 624 Hektor calculated by the polyhedron model method (Werner and Scheees 1997) with the shape data.

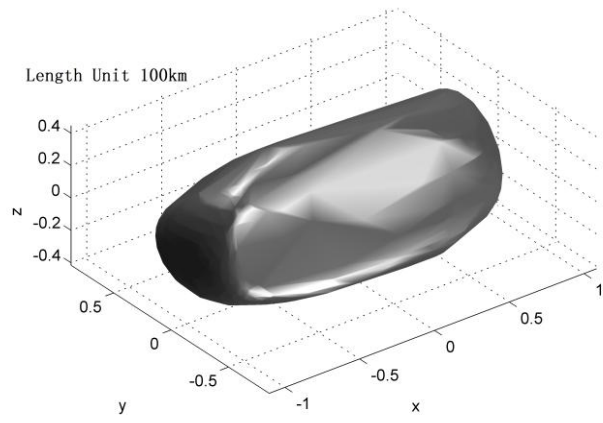

(a)

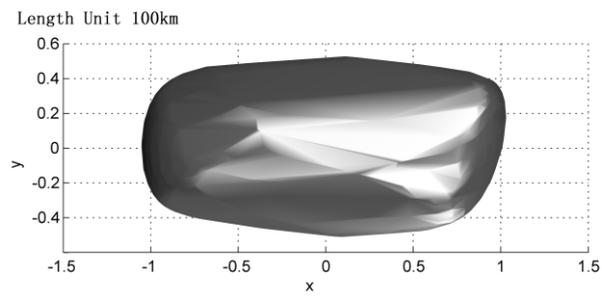

(b)

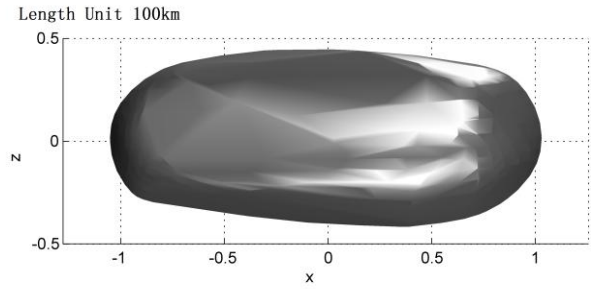

(c)

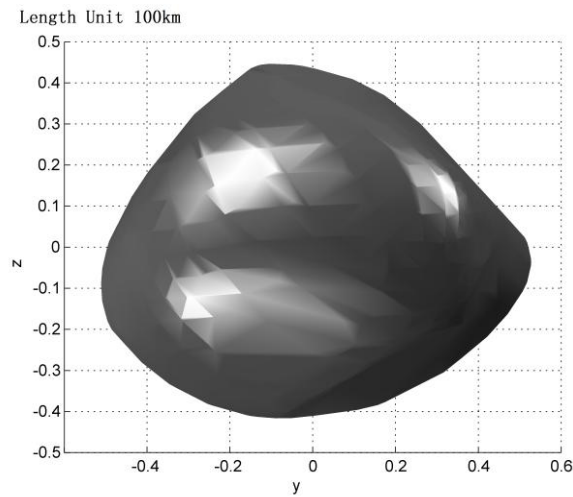

(d)

Figure 1. 3D convex shape of asteroid 624 Hektor using the polyhedron model, the shape model is calculated by 2038 faces and 1021 vertexes. (a) 3D view; (b) viewed in xy plane; (c) viewed in xz plane; (d) viewed in yz plane.

This paper has the following sections. Section 2 deals with the polyhedron model, gravitational potential, zero-velocity curves, as well as the equilibrium points of 624 Hektor with different density values. We also investigate the dynamical equations, effective potential, and eigenvalues of equilibrium points around an asteroid. The positions, eigenvalues, Jacobian, topological cases, and the stability of the equilibrium points around 624 Hektor are discussed. Section 3 covers the orbit in the potential of the primary of this asteroid. The periodic orbits and the orbit for the moonlet are calculated by considering the gravitational force acceleration caused by the irregular shape of the primary 624 Hektor. Section 4 gives a brief discussion of the results.

## 2 Gravitational Fields and Equilibrium Points

The estimated diameters of the moonlet and the primary are 12 km and 250 km, respectively. Thus the mass ratio of the moonlet and the primary is about $1.106 \times 10^{-4}$, and the distance of the barycenter of the system to the mass center of 624 Hektor is 0.1059 km. The gravitational influence of the moonlet on the equilibrium points and the orbits is small enough so that we can neglect any perturbation by the moonlet. For a particle about 150 km distance from the mass center of the asteroid, the ratio of Jupiter's gravitational force to the asteroid gravitational force is about $1.0 \times 10^{-5}$. The relative scales of other planetary perturbations are smaller than $1.0 \times 10^{-5}$. When the distance between the particle and the mass center of the asteroid decreases, the ratio of the Jupiter gravitational force to the asteroid gravitational force also decreases. By using the polyhedral model (Werner and Scheeres 1997) the dynamical equations of a particle orbiting in the potential of an asteroid can be written as Equation (1):

$$\ddot{\mathbf{r}} + 2\boldsymbol{\omega} \times \dot{\mathbf{r}} + \boldsymbol{\omega} \times (\boldsymbol{\omega} \times \mathbf{r}) + G\sigma \sum_{e \in edges} \mathbf{E}_e \cdot \mathbf{r}_e \cdot L_e - G\sigma \sum_{f \in faces} \mathbf{F}_f \cdot \mathbf{r}_f \cdot \omega_f = 0, \quad (1)$$

where the vectors are expressed in the body-fixed frame, $\mathbf{r} = (x, y, z)$ is the particle's position vector, $\boldsymbol{\omega}$ is the asteroid's rotational angular velocity relative to the inertial system, G=6.67×$10^{-11}$ $m^3kg^{-1}s^{-2}$ is the gravitational constant, $\sigma$ is the asteroid's density, $\mathbf{r}_e$ and $\mathbf{r}_f$ are vectors from the field points to any fixed points on edge and face, respectively; $\mathbf{E}_e$ is the dyad of edges related to two face- and edge-normal vectors, and $\mathbf{F}_f$ is the dyad calculated by the outer product of the outward-pointing surface normal vector with itself; $L_e$ is an integrated value of the general edge $e$ of face $f$ while $\omega_f$ is the signed solid angle subtended by the planar region.

The gravitational potential and effective potential can be calculated using Equation (2):

$$\begin{cases} U(\mathbf{r}) = -\frac{1}{2}G\sigma \sum_{e \in edges} \mathbf{r}_e \cdot \mathbf{E}_e \cdot \mathbf{r}_e \cdot L_e + \frac{1}{2}G\sigma \sum_{f \in faces} \mathbf{r}_f \cdot \mathbf{F}_f \cdot \mathbf{r}_f \cdot \omega_f \\ V(\mathbf{r}) = -\frac{1}{2}(\boldsymbol{\omega} \times \mathbf{r}) \cdot (\boldsymbol{\omega} \times \mathbf{r}) + U(\mathbf{r}) \end{cases} \quad (2)$$

The study of the equilibrium points can help us understand the dynamics and characteristics around the equilibrium points, and this study is essential when designing an orbit around the asteroid for the spacecraft. The locations of equilibrium points around the asteroid are critical points of the effective potential, thus we have the following Equation (3) to solve for the locations of equilibrium points:

$$\frac{\partial V(\mathbf{r})}{\partial \mathbf{r}} = 0. \quad (3)$$

The linearized dynamical equation relative to the equilibrium point can be written with Equation (4):

$$\frac{d^2}{dt^2}\begin{bmatrix}\xi \\ \eta \\ \zeta\end{bmatrix} + \begin{pmatrix} 0 & -2\omega_z & 2\omega_y \\ 2\omega_z & 0 & -2\omega_x \\ -2\omega_y & 2\omega_x & 0 \end{pmatrix} \cdot \frac{d}{dt}\begin{bmatrix}\xi \\ \eta \\ \zeta\end{bmatrix} + \begin{pmatrix} V_{xx} & V_{xy} & V_{xz} \\ V_{xy} & V_{yy} & V_{yz} \\ V_{xz} & V_{yz} & V_{zz} \end{pmatrix} \cdot \begin{bmatrix}\xi \\ \eta \\ \zeta\end{bmatrix} = 0, \quad (4)$$

where $\delta \mathbf{r} = [\xi \ \eta \ \zeta]^T = \mathbf{r}_E - \mathbf{r}$ is the particle's position vector relative to the equilibrium point, $\mathbf{r}_E$ is the position vector of the equilibrium point, $\boldsymbol{\omega} = \left[\omega_x, \omega_y, \omega_z\right]^T$, $V_{pq} = \left(\dfrac{\partial^2 V}{\partial p \partial q}\right)_E$ $(p, q = x, y, z)$. From the above equation one can see that the eigenvalues of the equilibrium point satisfy Eq. (5):

$$\begin{vmatrix} \lambda^2 + V_{xx} & -2\omega_z\lambda + V_{xy} & 2\omega_y\lambda + V_{xz} \\ 2\omega_z\lambda + V_{xy} & \lambda^2 + V_{yy} & -2\omega_x\lambda + V_{yz} \\ -2\omega_y\lambda + V_{xz} & 2\omega_x\lambda + V_{yz} & \lambda^2 + V_{zz} \end{vmatrix} = 0. \tag{5}$$

Here $\lambda$ is the eigenvalues.

The body-fixed frame is defined as the principal axis of inertia. The origin is the mass center of the asteroid. The x-axis parallels the smallest moment of inertia, while the z-axis parallels the largest moment of inertia. Assuming the asteroid rotates around its largest moment of inertia, then $\omega_z = |\boldsymbol{\omega}|$. In this case the linearized dynamical equations simplify to Eq. (6)

$$\dfrac{d^2}{dt^2}\begin{bmatrix}\xi\\\eta\\\zeta\end{bmatrix} + \begin{pmatrix}0 & -2\omega & 0\\ 2\omega & 0 & 0\\ 0 & 0 & 0\end{pmatrix}\cdot\dfrac{d}{dt}\begin{bmatrix}\xi\\\eta\\\zeta\end{bmatrix} + \begin{pmatrix}V_{xx} & V_{xy} & V_{xz}\\ V_{xy} & V_{yy} & V_{yz}\\ V_{xz} & V_{yz} & V_{zz}\end{pmatrix}\cdot\begin{bmatrix}\xi\\\eta\\\zeta\end{bmatrix} = 0. \tag{6}$$

The equation of eigenvalues reduces to Eq. (7):

$$\begin{vmatrix} \lambda^2 + V_{xx} & -2\omega\lambda + V_{xy} & V_{xz} \\ 2\omega\lambda + V_{xy} & \lambda^2 + V_{yy} & V_{yz} \\ V_{xz} & V_{yz} & \lambda^2 + V_{zz} \end{vmatrix} = 0. \tag{7}$$

The topological cases (Jiang et al. 2014; Wang et al. 2014) of the equilibrium points can be determined by the following distributions of eigenvalues, which are Case 1: $\pm i\beta_j \left(\beta_j \in \mathrm{R}, \beta_j > 0; j = 1, 2, 3\right)$; Case 2: $\pm\alpha_j\left(\alpha_j \in \mathrm{R}, \alpha_j > 0, j = 1\right)$, $\pm i\beta_j\left(\beta_j \in \mathrm{R}, \beta_j > 0; j = 1, 2\right)$; Case 3: $\pm\alpha_j\left(\alpha_j \in \mathrm{R}, \alpha_j > 0, j = 1, 2\right)$, $\pm i\beta_j\left(\beta_j \in \mathrm{R}, \beta_j > 0; j = 1\right)$; Case 4a: $\pm\alpha_j\left(\alpha_j \in \mathrm{R}, \alpha_j > 0, j = 1\right)$, $\pm\sigma \pm i\tau\left(\sigma, \tau \in \mathrm{R}; \sigma, \tau > 0\right)$; Case 4b: $\pm\alpha_j\left(\alpha_j \in \mathrm{R}, \alpha_j > 0, j = 1, 2, 3\right)$; Case 5:

$\pm i\beta_j \left(\beta_j \in \mathrm{R}, \beta_j > 0, j = 1\right)$, $\pm\sigma \pm i\tau \left(\sigma, \tau \in \mathrm{R}; \sigma, \tau > 0\right)$. The distribution of the eigenvalues on the complex plane are shown in Figure 2. For Case 1 the equilibrium point is linearly stable. For Case 2 and Case 5, the equilibrium points are unstable.

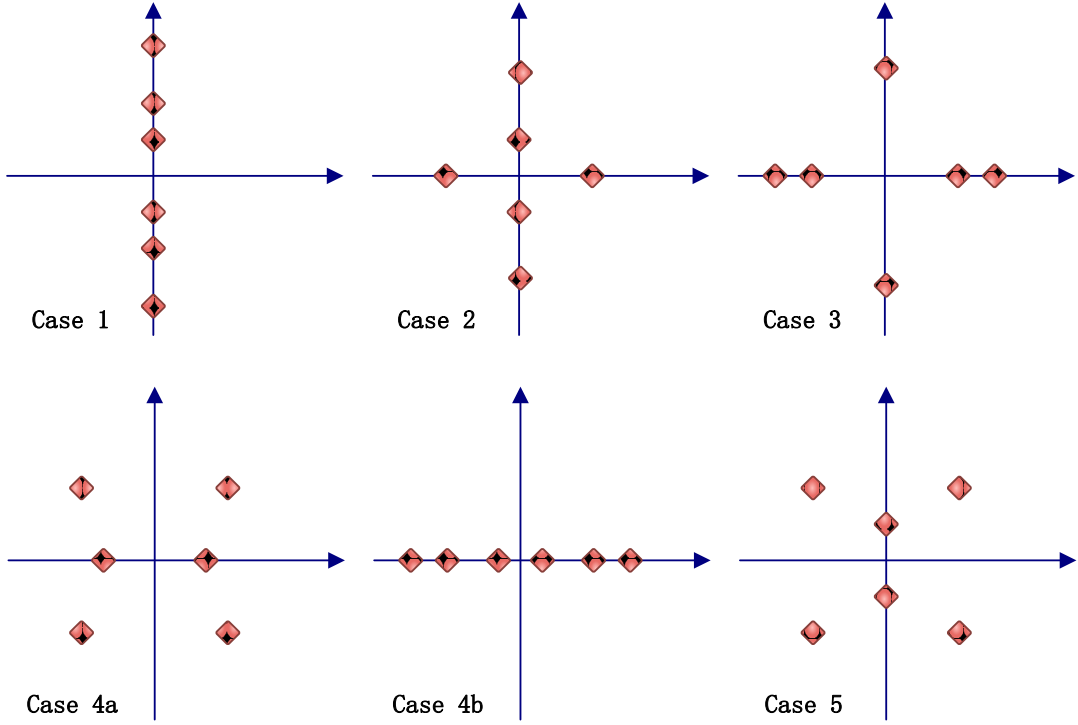

Figure 2. The distribution of the eigenvalues on the complex plane for the different topological cases, the red diamond symbols mean the locations of the eigenvalues.

The Jacobian integral $H$ (Scheeres et al. 1996) is given by Equation (8):

$$H = \frac{1}{2}\dot{\mathbf{r}} \cdot \dot{\mathbf{r}} - \frac{1}{2}(\boldsymbol{\omega} \times \mathbf{r})(\boldsymbol{\omega} \times \mathbf{r}) + U(\mathbf{r}).\tag{8}$$

Because the dynamical system of Eq. (1) is a conservative system, the Jacobian integral is a constant. There are three different density values presented in previous work. Figure 3 presents gravitational parameters of asteroid 624 Hektor with the different density values ρ=2.43 g·cm$^{-3}$, 1.63 g·cm$^{-3}$, and 1.0 g·cm$^{-3}$ respectively.

## 2.1 Gravitation Parameters with Density 2.43 g·cm$^{-3}$

Descamps (2014) derived the bulk density of 624 Hektor to be 2.43 g·cm$^{-3}$. Here we use this parameter of the density to calculate the effective potential and equilibrium points of 624 Hektor. Figure 3 presents the gravitational parameters of 624 Hektor with the density as ρ=2.43 g·cm$^{-3}$. The structure of the effective potential in the xy, yz, and zx planes are different. There are five equilibrium points in the gravitational potential of 624 Hektor; one is inside, and the other four are outside. Table 1 gives the positions and Jacobian of the equilibrium points when the density is ρ=2.43 g·cm$^{-3}$. From Table 1 one can see that all of these five equilibrium points are not in the equatorial plane of 624 Hektor; i.e. all of them are out-of-plane equilibrium points. E5 is near the mass center of the body inside the body of the asteroid. The other equilibrium points E1-E4 are outside. E1 and E3 are near the x axis while E2 and E4 are near the y axis. In other words, E1 and E3 are near the axis of smallest moment of inertia while E2 and E4 are near the axis of intermediate moment of inertia. We refer to E1 and E3 as the equilibrium points associated with the smallest moment of inertia, E2 and E4 as the equilibrium points associated with the intermediate moment of inertia, and E5 associated with the mass center of the body. The equilibrium point E5 has the maximum value of the Jacobian. The equilibrium points E4 and E2, which are associated with the intermediate moment of inertia, have the minimum value of the Jacobian. And the equilibrium points E3 and E1, which are associated with the smallest moment of inertia, have the intermediate value of the Jacobian.

Table 2 presents eigenvalues of the equilibrium points around 624 Hektor with the density equal to 2.43 g·cm$^{-3}$, and Table 3 presents the topological cases and stability of the equilibrium points. The topological case of the equilibrium point E5 belongs to Case 1, thus one can conclude that the equilibrium point E5 is linearly stable. From the positive definite of the Hessian matrix of the equilibrium point E5, one can further conclude that the equilibrium point E5 is stable.

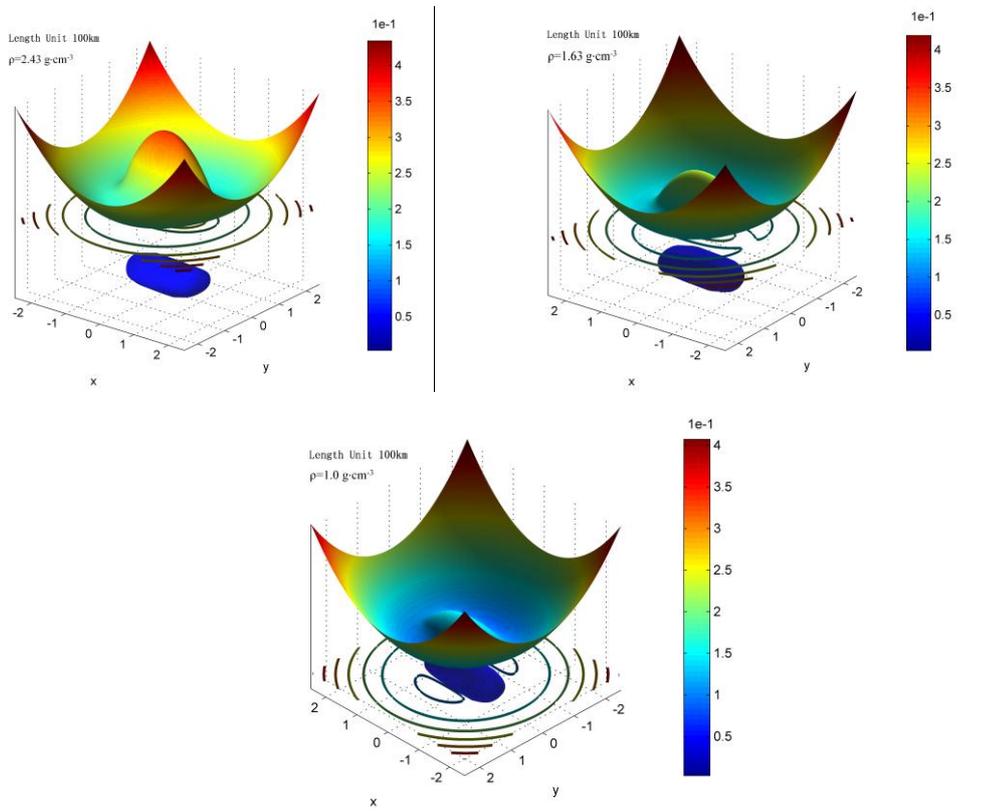

(a)

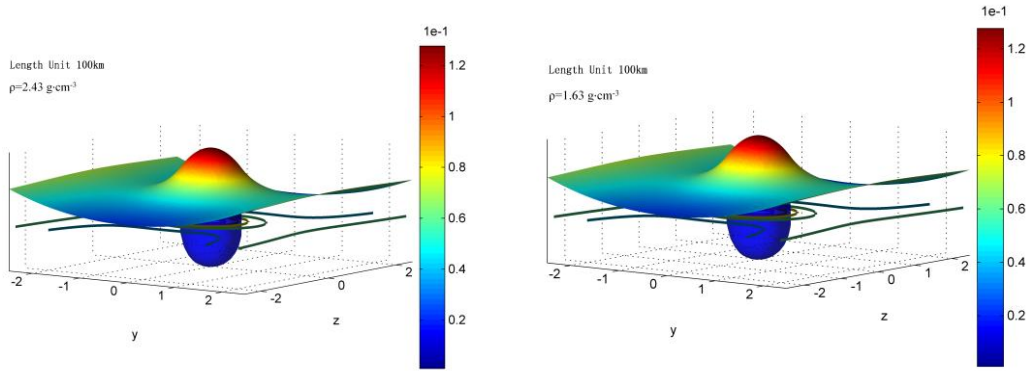

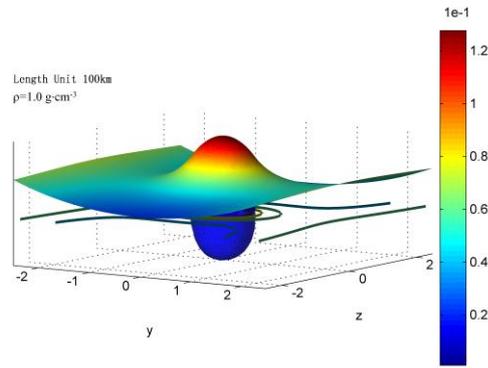

(b)

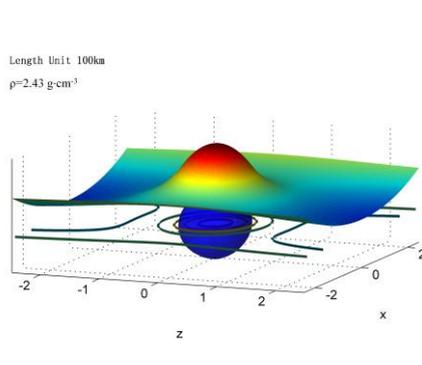
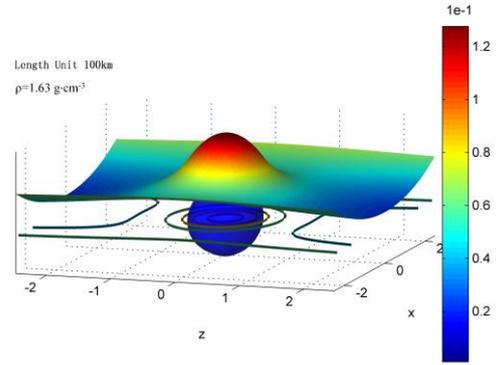

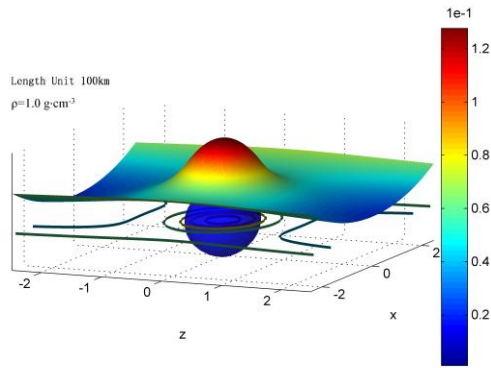

(c)

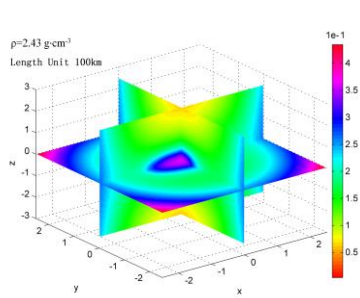
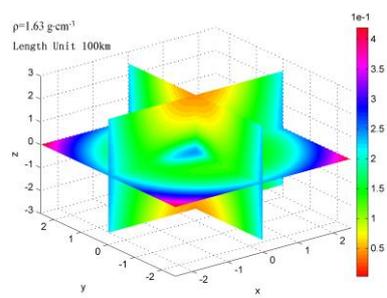

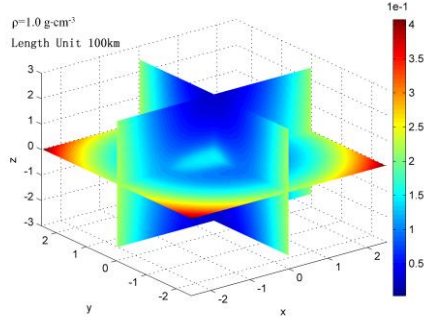

(d)

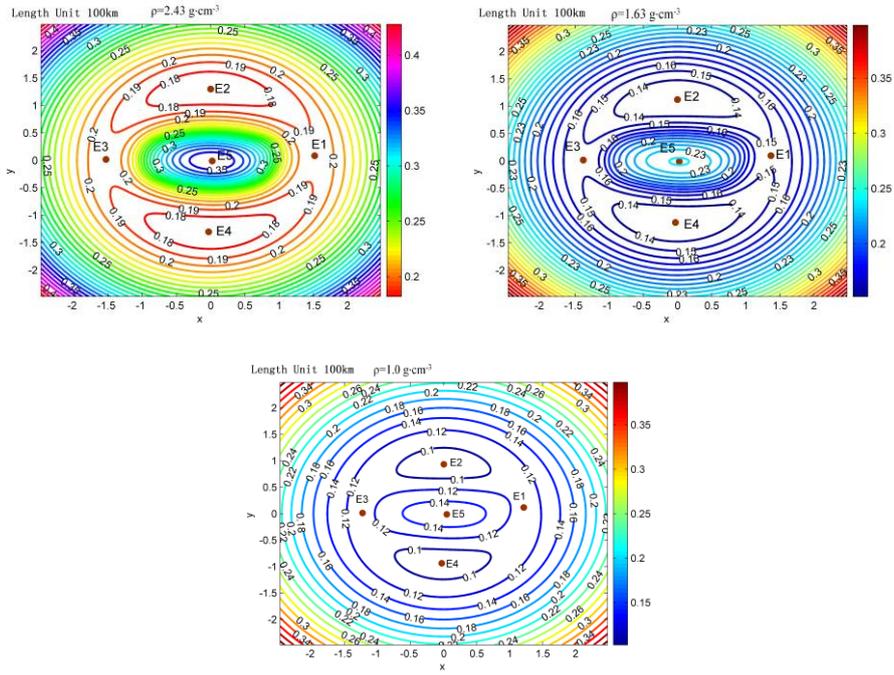

(e)

Figure 3. Gravitational parameters of asteroid 624 Hektor with the different density values ρ=2.43 g·cm$^{-3}$, 1.63 g·cm$^{-3}$, and 1.0 g·cm$^{-3}$, respectively. The unit of the effective potential is $10^4$ m$^2$·s$^{-2}$. The length unit in (a), (b), and (c) is 100 km. (a) Effective potential in the xy plane of the body-fixed frame of the asteroid; (b) Effective potential in the yz plane of the body-fixed frame of the asteroid; (c) Effective potential in the zx plane of the body-fixed frame of the asteroid; (d) 3D plot of the effective potential; (e) The contour plot of the effective potential in the xy plane and the projection of equilibrium points.

Table 1 Positions and Jacobian of the equilibrium points around 624 Hektor with the density 2.43 g·cm$^{-3}$

| Equilibrium Points | x (km) | y (km) | z (km) | Altitude (km) | Jacobian ($10^4$m$^2$s$^{-2}$) |
| --- | --- | --- | --- | --- | --- |
| E1 | 151.883 | 8.49373 | 0.587014 | 49.8214 | 0.192206 |
| E2 | -0.161827 | 129.870 | -0.461023 | 77.8709 | 0.172092 |
| E3 | -152.994 | 1.66306 | 0.0447414 | 47.9530 | 0.192866 |
| E4 | -3.14642 | -130.242 | -0.601781 | 80.2814 | 0.172393 |
| E5 | 2.13020 | -0.969565 | -0.0792624 | | 0.359340 |

Table 2 Eigenvalues of the equilibrium points around 624 Hektor with the density 2.43 g·cm$^{-3}$

| Equilibrium Points (×10$^{-3}$s$^{-1}$) | $\lambda_1$ | $\lambda_2$ | $\lambda_3$ | $\lambda_4$ | $\lambda_5$ | $\lambda_6$ |
|---|---|---|---|---|---|---|
| E1 | 0.302938i | -0.302938i | 0.298380i | -0.298380i | 0.231548 | -0.231548 |
| E2 | 0.256737i | -0.256737i | 0.115225+0.209556i | 0.115225-0.209556i | -0.115225+0.209556i | -0.115225-0.209556i |
| E3 | 0.307942i | -0.307942i | 0.301666i | -0.301666i | 0.242163 | -0.242163 |
| E4 | 0.255683i | -0.255683i | 0.109262+0.206991i | 0.109262-0.206991i | -0.109262+0.206991i | -0.109262-0.206991i |
| E5 | 1.02580i | -1.02580i | 0.995960i | -0.995960i | 0.347936i | -0.347936i |

Table 3 The topological cases and stability of the equilibrium points around 624 Hektor with the density 2.43 g·cm$^{-3}$. S: stable; U: unstable; P: positive definite; N: non-positive definite; Index of inertia: positive/ negative index of inertia

| Equilibrium Points | Topological Case | Stability | $\nabla^2 V$ | Index of Inertia |
|---|---|---|---|---|
| E1 | 2 | U | N | 2/1 |
| E2 | 5 | U | N | 1/2 |
| E3 | 2 | U | N | 2/1 |
| E4 | 5 | U | N | 1/2 |
| E5 | 1 | S | P | 3/0 |

## 2.2 Gravitation Parameters with the Density 1.63 g·cm$^{-3}$

Carry (2012) reported the masses, density, and diameter of 287 objects; the density of 624 Hektor was estimated to be 1.63 g·cm$^{-3}$. The number of equilibrium points is also five. The gravitational parameters of asteroid 624 Hektor with the density ρ=1.63 g·cm$^{-3}$ can be seen in Figure 3. Table 4 presents positions and Jacobian of the equilibrium points while Table 5 presents eigenvalues. Table 6 shows the topological cases and stability of the equilibrium points. The topological cases and stability of the equilibrium points in this case are the same as those when the density is ρ=2.43 g·cm$^{-3}$. The five equilibrium points are all out-of-plane equilibrium points, which are the same as those when ρ=2.43 g·cm$^{-3}$. For the equilibrium points E1 and E3 which are associated with the smallest moment of inertia, the components of the position vectors on the smallest moment of inertia are smaller than the components when the density is ρ=2.43 g·cm$^{-3}$, and the components of the position vectors on the intermediate moment of inertia are also smaller than the components when the density ρ=2.43 g·cm$^{-3}$. However, the component of the position vector of the equilibrium point E1 onto the intermediate moment of inertia is larger than the component when the density is ρ=2.43 g·cm$^{-3}$.

For the equilibrium points E2 and E4 which are associated with the intermediate moment of inertia, the components of the position vector on the intermediate moment of inertia are also smaller than the components when the density is ρ=2.43 g·cm$^{-3}$. However, for the equilibrium point E2 the absolute value of the component of the position vector onto the smallest moment of inertia is larger than the component when the density is ρ=2.43 g·cm$^{-3}$.

Considering the absolute value of the component of the position vector on the largest moment of inertia, the values of equilibrium points E1, E2, and E4 when the density is ρ=1.63 g·cm$^{-3}$ are larger than the values when the density is ρ=2.43 g·cm$^{-3}$. When considering the absolute value of the components of the position vector of the equilibrium point E5, the value for the density with ρ=1.63 g·cm$^{-3}$ is larger than the value for the density with ρ=2.43 g·cm$^{-3}$.

In general the norm values of the position vectors of the outer equilibrium points on the smallest moment of inertia when the density is ρ=1.63 g·cm$^{-3}$ are smaller than the norm values when the density is ρ=2.43 g·cm$^{-3}$.

Table 4 Positions and Jacobian of the equilibrium points around 624 Hektor with the density 1.63 g·cm$^{-3}$

| Equilibrium Points | x (km) | y (km) | z (km) | Altitude (km) | Jacobian ($10^4$m$^2$s$^{-2}$) |
|---|---|---|---|---|---|
| E1 | 136.764 | 9.68405 | 0.765553 | 34.8086 | 0.150627 |
| E2 | -0.0492236 | 112.108 | -0.599798 | 60.1096 | 0.130591 |
| E3 | -138.164 | 1.45284 | 0.0278511 | 33.1216 | 0.151366 |
| E4 | -3.15940 | -112.527 | -0.804599 | 62.5742 | 0.130903 |
| E5 | 2.61313 | -1.02397 | -0.0855378 |  | 0.241045 |

Table 5 Eigenvalues of the equilibrium points around 624 Hektor with the density 1.63 g·cm$^{-3}$

| Equilibrium Points (× $10^{-3}$s$^{-1}$) | $\lambda_1$ | $\lambda_2$ | $\lambda_3$ | $\lambda_4$ | $\lambda_5$ | $\lambda_6$ |
|---|---|---|---|---|---|---|
| E1 | 0.314086i | -0.314086i | 0.311102i | -0.311102i | 0.261240 | -0.261240 |
| E2 | 0.258520i | -0.258520i | 0.127639+0.215568i | 0.127639-0.215568i | -0.127639+0.215568i | -0.127639-0.215568i |
| E3 | 0.320620i | -0.320620i | 0.315223i | -0.315223i | 0.273815 | -0.273815 |
| E4 | 0.256805i | -0.256805i | 0.121326+0.212931i | 0.121326-0.212931i | -0.121326+0.212931i | -0.121326-0.212931i |
| E5 | 0.877748i | -0.877748i | 0.815814i | -0.815814i | 0.241448i | -0.241448i |

Table 6 The topological cases and stability of the equilibrium points around 624 Hektor with the density 1.63 g·cm$^{-3}$. S: stable; U: unstable; P: positive definite; N: non-positive definite; Index of inertia: positive/ negative index of inertia

| Equilibrium Points | Topological Case | Stability | $\nabla^2 V$ | Index of Inertia |
|---|---|---|---|---|
| E1 | 2 | U | N | 2/1 |
| E2 | 5 | U | N | 1/2 |
| E3 | 2 | U | N | 2/1 |
| E4 | 5 | U | N | 1/2 |
| E5 | 1 | S | P | 3/0 |

## 2.3 Gravitation Parameters with the Density equal to 1.0 g·cm$^{-3}$

Marchis et al. (2014) listed the parameters of asteroid 624 Hektor, and the density from this work is ρ=1.0 g·cm$^{-3}$. The gravitational parameters of asteroid 624 Hektor with the density ρ=1.0 g·cm$^{-3}$ can be seen in Figure 3. Table 7 presents the positions and Jacobian of the equilibrium points in the potential of 624 Hektor, and Table 8 presents eigenvalues of the equilibrium points.

Table 9 illustrates the topological cases, stability, positive definite or not, and positive/ negative index of inertia of the equilibrium points. Comparing Table 9 with Table 3 and Table 6, one can conclude that the topological cases and stability of the equilibrium points for these three cases, ρ=1.0 g·cm$^{-3}$, 1.63 g·cm$^{-3}$, and 2.43 g·cm$^{-3}$ are the same. Although all the five equilibrium points remain out-of-plane, the distances of the equilibrium points relative to the equatorial plane are different from the above two cases with ρ=1.63 g·cm$^{-3}$ and 2.43 g·cm$^{-3}$. The distances of the outer equilibrium points E1-E4 relative to the mass center of the asteroid for ρ=1.0 g·cm$^{-3}$ are smaller than for ρ=1.63 g·cm$^{-3}$ and 2.43 g·cm$^{-3}$.

For the equilibrium points E1 and E3 which are associated with the smallest moment of inertia, we consider the components of the position vectors of the equilibrium points on the smallest moment of inertia; if the density is ρ=2.43 g·cm$^{-3}$, the component is the largest one; if the density is ρ=1.0 g·cm$^{-3}$, the component is the smallest one. Similarly, for the equilibrium points E2 and E4 which are associated with the intermediate moment of inertia, we consider the components of the position vectors of the equilibrium points onto the intermediate moment of inertia; if the density is ρ=2.43 g·cm$^{-3}$, the component is the largest one; if the density is ρ=1.0 g·cm$^{-3}$, the component is the smallest one.

For the equilibrium point E5 which is associated with the mass center of the body we consider the distance of the equilibrium point and the mass center of the body. When the density is ρ=1.0 g·cm$^{-3}$, the value of the distance is the largest one; when the density is ρ=2.43 g·cm$^{-3}$, the value of the distance is the smallest one. Additionally, considering the absolute value of the components of the position vector of the equilibrium point E5; the value is the largest one when the density is ρ=1.0 g·cm$^{-3}$, while the value is the smallest one when the density is ρ=2.43 g·cm$^{-3}$.

Now we consider the norm value of the position vector of the outer equilibrium points on the smallest moment of inertia; if the density is ρ=1.0 g·cm$^{-3}$, the value is the smallest one; if the density is ρ=2.43 g·cm$^{-3}$, the value is the largest one.

Table 7 Positions and Jacobian of the equilibrium points around 624 Hektor with the density 1.0 g·cm$^{-3}$

| Equilibrium Points | x (km) | y (km) | z (km) | Altitude (km) | Jacobian ($10^4$m$^2$s$^{-2}$) |
|---|---|---|---|---|---|
| E1 | 121.325 | 11.7195 | 1.02686 | 19.5940 | 0.112761 |
| E2 | 0.0342004 | 93.2722 | -0.835230 | 41.2759 | 0.0928577 |
| E3 | -123.128 | 1.20486 | 0.0146154 | 18.0839 | 0.113585 |

| E4 | -3.12573 | -93.7164 | -1.15650 | 43.7756 | 0.0931712 |
| E5 | 4.58001 | -1.16954 | -0.110279 | \ | 0.147897 |

Table 8 Eigenvalues of the equilibrium points around 624 Hektor with the density 1.0 g·cm$^{-3}$

| Equilibrium Points (× $10^{-3}s^{-1}$) | $\lambda_1$ | $\lambda_2$ | $\lambda_3$ | $\lambda_4$ | $\lambda_5$ | $\lambda_6$ |
| --- | --- | --- | --- | --- | --- | --- |
| E1 | 0.337176i | -0.337176i | 0.324412i | -0.324412i | 0.302891 | -0.302891 |
| E2 | 0.262029i | -0.262029i | 0.141692+0.222128i | 0.141692-0.222128i | -0.141692+0.222128i | -0.141692-0.222128i |
| E3 | 0.344678i | -0.344678i | 0.329035i | -0.329035i | 0.316036 | -0.316036 |
| E4 | 0.258769i | -0.258769i | 0.134573+0.219598i | 0.134573-0.219598i | -0.134573+0.219598i | -0.134573-0.219598i |
| E5 | 0.735719i | -0.735719i | 0.638779i | -0.638779i | 0.128765i | -0.128765i |

Table 9 The topological cases and stability of the equilibrium points around 624 Hektor with the density 1.0 g·cm$^{-3}$. S: stable; U: unstable; P: positive definite; N: non-positive definite; Index of inertia: positive/ negative index of inertia

| Equilibrium Points | Topological Case | Stability | $\nabla^2 V$ | Index of Inertia |
| --- | --- | --- | --- | --- |
| E1 | 2 | U | N | 2/1 |
| E2 | 5 | U | N | 1/2 |
| E3 | 2 | U | N | 2/1 |
| E4 | 5 | U | N | 1/2 |
| E5 | 1 | S | P | 3/0 |

## 2.4 Discussions of the Three Cases

Sections 2.1-2.3 investigated the positions, Jacobian, eigenvalues, topological cases, stability, and the Hessian matrix of the equilibrium points with different density values. From the results one can conclude that the topological cases and the stability of the equilibrium points with different density values are the same. The positive definite or not and the positive/ negative index of inertia of the Hessian matrix are also the same for the three cases of the density values.

The distances between the outer equilibrium points (including E1-E4) and the mass center decrease when the density value of the asteroid 624 Hektor decreases. For the case of density equal to $\rho=2.43$ g·cm$^{-3}$, the lengths of the position vectors of the outer equilibrium points are the largest. For the case of density equal to $\rho=1.0$ g·cm$^{-3}$, the lengths of the position vectors of the outside equilibrium points are the smallest. The distance of the inner equilibrium point E5 and the mass center increases when the density value of the asteroid 624 Hektor decreases.

Now the Trojan asteroid 624 Hektor does not have precise data for a shape model and density. The location, stability, and topological cases of equilibrium points are related to the shape model, density, and rotational speed of the body. The initial values and period of the periodic orbits are also related to these body parameters. The contact binary asteroids, such as 216 Kleopatra, 2063 Bacchus, 4769 Castalia, and 25143 Itokawa have different shapes. Asteroid 216 Kleopatra obviously has a double-lobed structure, and looks like a dumbbell. For others, including 2063 Bacchus, 4769 Castalia, and 25143 Itokawa, the double-lobed structure looks obscure. The polyhedron shape data of 624 Hektor shows the shape is convex. Here we change the shape of 624 Hektor to the double-lobed structure, and calculate the effective potential around the asteroid to analyze the influence of the error of the shape to the error of the position error of equilibrium points.

The method to generate the double-lobed structure is given by Equation (9):

$$\text{If } |x| < L, \begin{cases} y_{new} = y_{min} + (y - y_{min})\dfrac{|x|}{L} \\ z_{new} = z_{min} + (z - z_{min})\dfrac{|x|}{L} \end{cases} \quad (9)$$

where $(x, y, z)$ are the surface points, $r = \sqrt{x^2 + y^2 + z^2}$, $L = 35\,\text{km}$, $y_{min} = z_{min} = 20\,\text{km}$ give the radius of the neck of the double-lobed structure, $y_{new}$ and $z_{new}$ are the new coordinate components of the surface points.

Then the new shape of the asteroid is concave and has a double-lobed structure. Figure 4 shows the shape change to the double-lobed structure. The x components of E1 and E3 are 108.864 km and -111.446 km, respectively; the y components of E2 and E4 are 78.1089 km and -78.3021 km, respectively. For the equilibrium points E1 and E3, the relative errors of the double-lobed structure referred to the above model are 28.3% and 27.2%, respectively; and for the equilibrium points E2 and E4, the relative errors are 39.89% and 39.88%, respectively. The topological case and stability of the equilibrium points remain unchanged.

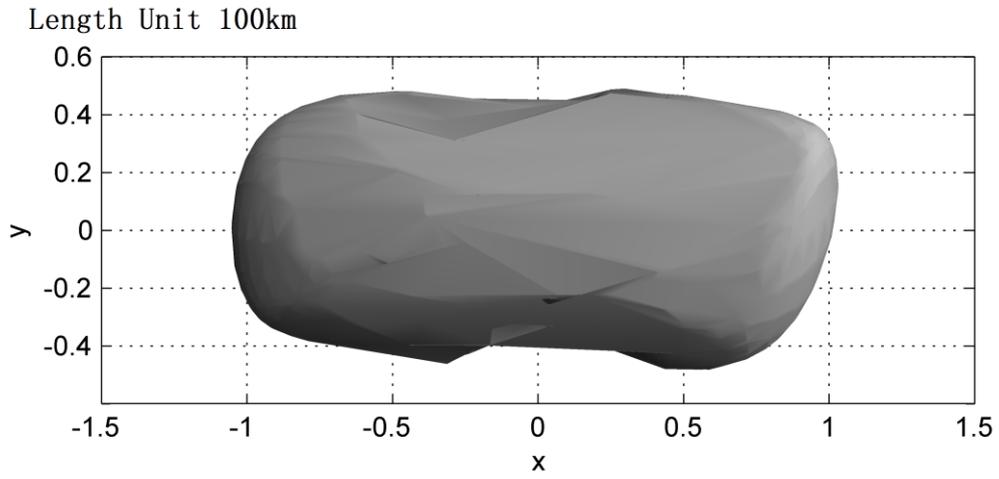

(a)

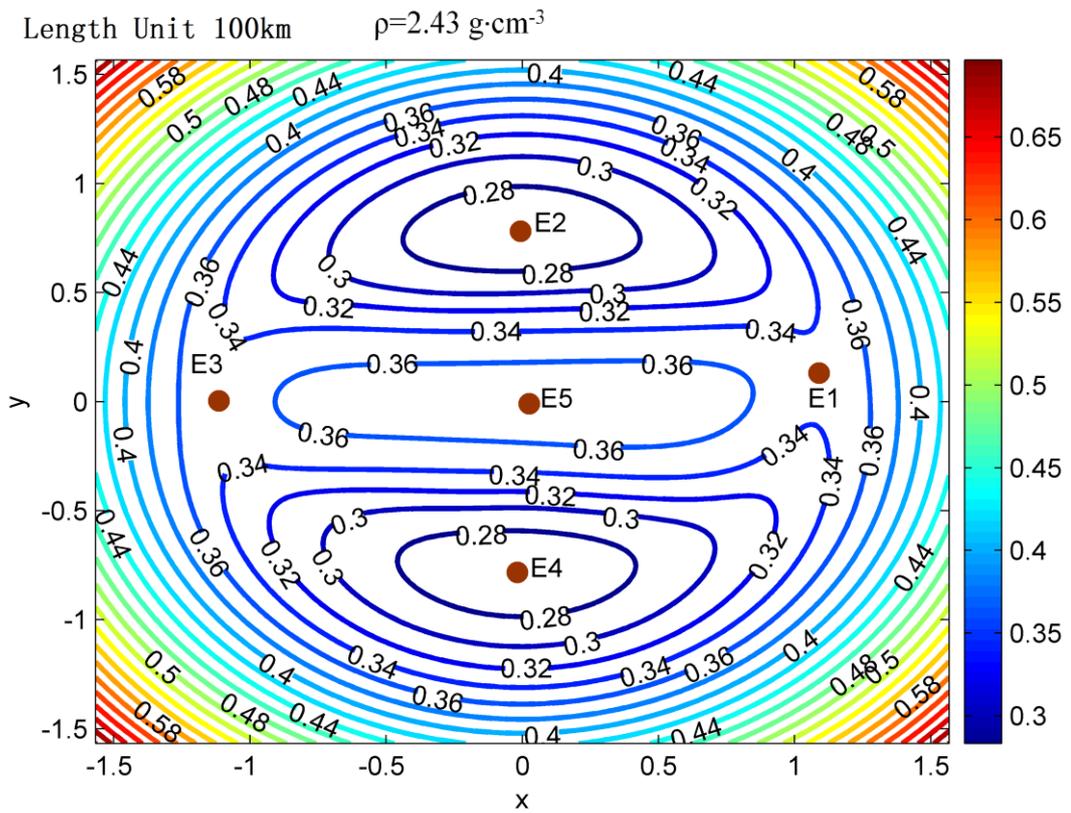

(b)

Figure 4 The shape change to the double-lobed structure. (a) The shape viewed in xy plane; (b) The contour plot of the effective potential in the xy plane and the projection of equilibrium points.

**3 Orbits around 624 Hektor**

The binary asteroid system 624 Hektor and Skamandrios is a large-size-ratio binary system. The equivalent diameter of the primary 624 Hektor is 250 km, while the estimated diameter of the moonlet Skamandrios is 12 km (Marchis et al. 2014). The dynamical J2 of 624 Hektor is calculated to be 0.15 (Marchis et al. 2014). The study of the orbits in the gravitational potential of the primary is useful when analyzing the complicated dynamical behaviors of the large-size-ratio binary Trojan asteroids. In this section we first study periodic orbits around the primary 624 Hektor which are near the surface. Then we calculate the orbits of the moonlet in the gravitational potential of the primary 624 Hektor. The density of 624 Hektor we use here is 1.0 g·cm$^{-3}$. If a periodic orbit exists in the gravitational potential of a uniformly rotating body with an initial density, there also exists a periodic orbit in the same positions with the same distribution of Floquet multipliers, although the initial velocity is different (Liu et al. 2011). Thus if we use the density value 1.0 g·cm$^{-3}$ to calculate the orbits and get results, the qualitative conclusions are also suitable for other density values.

To study the orbits more accurately we do not use the spherical harmonic expansion of 624 Hektor. The model of the spherical harmonic expansion (for instance, 10×10 orders) for a highly irregular shaped body has low precision, and may diverge at some points (Elipe and Riaguas 2003). In addition if one wants the spherical harmonic expansion to be accurate enough when the particle orbits outside the reference radius of the primary, the number of orders should be very large. Thus the orbits here are computed by considering the complete gravitational potential caused by the 3D irregular shape of the primary. In section 3.1 we investigate the periodic orbits near the surface of the primary 624 Hektor. In section 3.2 the orbit of the moonlet Skamandrios in the gravitational potential of the primary 624 Hektor is analyzed.

**3.1 Periodic Orbits near the Surface of 624 Hektor**

For the periodic orbit in the potential of a uniformly rotating asteroid we define the state transition matrix $\Phi(\cdot)$ as Equation (10)

$$\Phi(t) = \int_0^t \frac{\partial \mathbf{f}}{\partial \mathbf{z}}(p(\tau)) d\tau, \quad (10)$$

where $p$ denotes the periodic orbit, $\mathbf{z}(t) = \mathbf{f}(t, \mathbf{z}_0)$ represents the orbit which satisfies $\mathbf{f}(0, \mathbf{z}_0) = \mathbf{z}_0$, $t$ is the time. The orbit can be also written as a flow of the dynamical system as $\mathbf{f}^t : \mathbf{z} \to \mathbf{f}(t, \mathbf{z})$. A periodic orbit with period $T$ means $\Phi(0) = \Phi(T)$. Define the monodromy matrix of the periodic orbit as $M = \Phi(T)$. The stability of the periodic orbit depends on the distribution of eigenvalues of the monodromy matrix. The eigenvalues of the monodromy matrix are also called Floquet multipliers. The Floquet multipliers of a periodic orbit in the potential of a uniformly rotating asteroid have the form 1, $-1$, $\mathrm{sgn}(\alpha) e^{\pm \alpha} \left( \alpha \in \mathrm{R}, |\alpha| \in (0,1) \right)$, $e^{\pm i\beta} \left( \beta \in (0, \pi) \right)$, and $e^{\pm \sigma \pm i\tau} \left( \sigma, \tau \in \mathrm{R}; \sigma > 0, \tau \in (0, \pi) \right)$. Here

$\text{sgn}(\alpha) = \begin{cases} 1 & (\text{if } \alpha > 0) \\ -1 & (\text{if } \alpha < 0) \end{cases}$. Because the system is a Hamiltonian system, there are at least two Floquet multipliers equal to +1 for a periodic orbit. Figure 5 and Figure 6 present two different periodic orbits near the surface of the primary 624 Hektor.

The positions of these two periodic orbits are different, but the shapes look similar. From the distribution of Floquet multipliers in the complex plane one can conclude that the two periodic orbits are unstable; the Floquet multipliers are in the form of 1, 1, $e^{\pm i\beta}$ $(\beta \in (0, \pi))$, and $e^{\pm\sigma \pm i\tau}$ $(\sigma, \tau \in \mathbb{R};\ \sigma > 0, \tau \in (0, \pi)\ )$. There are two Floquet multipliers on the unit circle except $\pm 1$, and two Floquet multipliers on the x axis except $\pm 1$. A periodic orbit can be continued to several periodic orbits, these periodic orbits have the same stability before the bifurcation occurs. There also exist periodic orbits that are triply-periodic, which repeat every 3 orbits. Figure 7 presents a periodic orbit which is triply-periodic. This periodic orbit is 2:1 resonant, i.e. the ratio of the period of the orbit and the rotational period of the asteroid is 2:1. The periodic orbit is also stable.

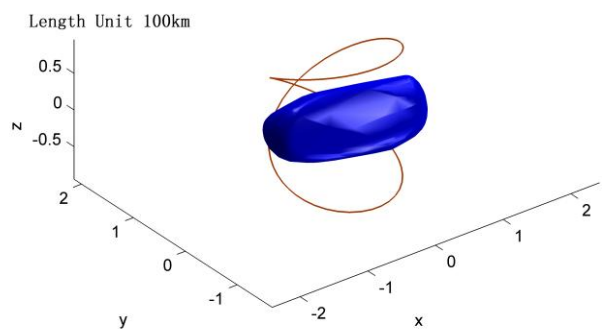

(a)

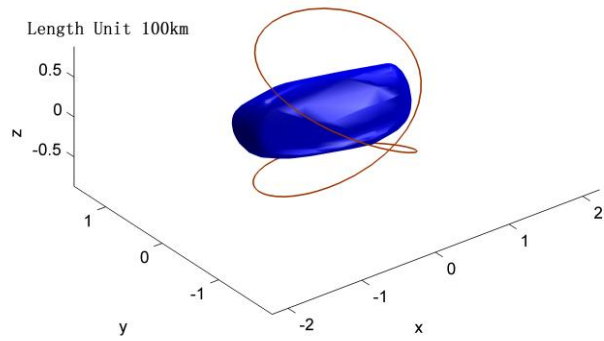

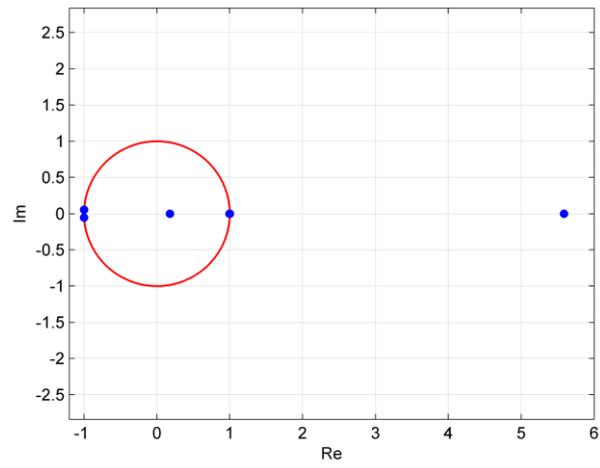

(b)

Figure 5. The first periodic orbit near the surface of the primary 624 Hektor. (a) The geometry of the orbit in the body-fixed frame of 624 Hektor; (b) The distribution of Floquet multipliers in the complex plane.

(a)

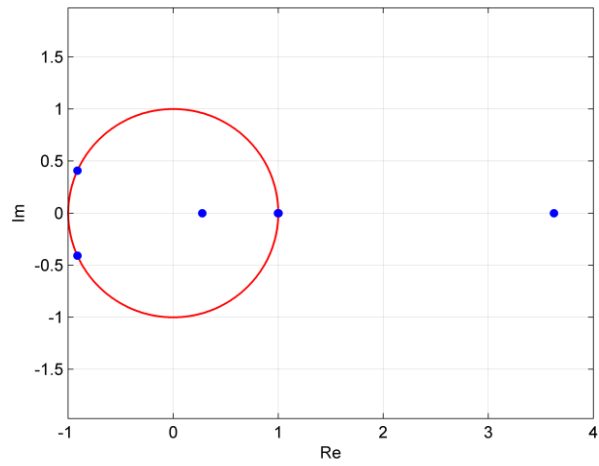

Figure 6. The second periodic orbit near the surface of the primary 624 Hektor. (a) The geometry of the orbit in the body-fixed frame of 624 Hektor; (b) The distribution of Floquet multipliers in the complex plane.

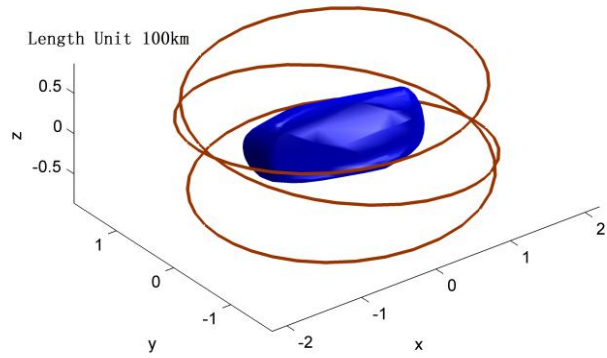

(a)

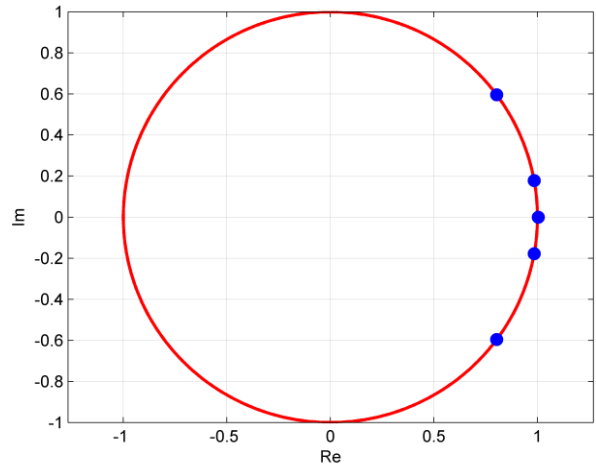

(b)

Figure 7. The 2:1 resonant periodic orbit which repeats every 3 orbits near the surface of the primary 624 Hektor. (a) The geometry of the orbit in the body-fixed frame of 624 Hektor; (b) The distribution of Floquet multipliers in the complex plane.

**3.2 The Simulation of the Orbit for the Moonlet in the Gravitational Potential of the Primary Caused by its 3D Irregular Shape**

In this section we simulate the orbit for the moonlet Skamandrios in the gravitational potential of the primary 624 Hektor. The gravitational potential of 624 Hektor caused by the 3D irregular shape is calculated. The distance from Skamandrios to the mass center of (624) Hektor is about 623.5 km (Marchis et al. 2014). The reference coordinate system is the body-fixed frame; we set the initial position of the moonlet at the x axis with the distance 623.5 km from the primary's mass center, and the velocity of the moonlet is in the xy plane and perpendicular to the x axis. Using this coordinate system one can see the motion of the moonlet relative to the primary. Figure 8 illustrates the results of the simulation of the orbit of the moonlet in the potential of 624 Hektor. The integral time is 28.8375 d. The integrator used here is the variable step-size 7/8 order Runge-Kutta, the relative error of the integrator is set to be $1.0 \times 10^{-13}$, the maximum and minimum step-sizes of the integrator are 689.54 s and 443.569 s, respectively. The orbit is near the primary's equatorial plane, and the orbit period is larger than the rotational period of the primary. This implies that if one stays on the surface of the primary, one will see that the particle on the orbit rotates in a westerly direction. If one stays in inertia space, one will find that the particle on the orbit rotates slower than the primary. From Figure 8 one can conclude that the orbit continues to be stable after 28.8375 d.

Jiang et al. (2015) investigated the periodic orbit families in the gravitational potential of the primary for the large-size-ratio triple asteroid 216 Kleopatra in order to analyze the stable regions. The periodic orbit families are near the equatorial plane of the primary, nearly circular, and stable. The two moonlets, Alexhelios and Cleoselene, of this large-size-ratio triple asteroid 216 Kleopatra move near the primary's equatorial plane (Descamps et al. 2011; Jiang et al. 2016). The result can help us to understand the orbit stability of the moonlets Alexhelios and Cleoselene. Lan et al. (2017) investigate the periodic orbit families to analyze the stable and unstable regions in the gravitational potential of 4 Vesta, 216 Kleopatra, 243 Ida, and 433 Eros. The literature uses the stability of periodic orbit families to analyze the motion stability of general orbits. However, the general orbit is not the periodic orbit, although the general orbit includes the periodic one. Here we analyze the motion stability of general orbits and the orbits of the moonlets using the general orbit itself.

The geometrical shapes of the orbit relative to the body-fixed frame and the orbit relative to the inertia system are different. The orbit relative to the body-fixed frame has a larger amplitude. The mechanical energy of the orbit varies periodically. The period of the variation of the orbit's mechanical energy and the period of the primary are the same. The Jacobi integral of the orbit is conservative, which equals -1693.4265 J·kg$^{-1}$.

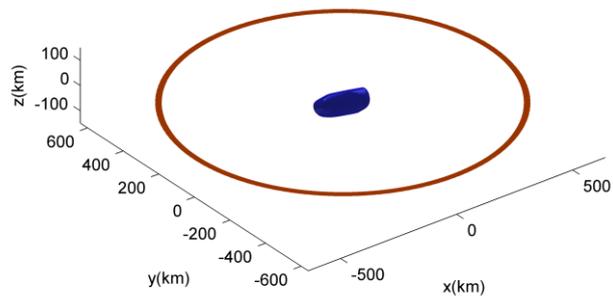

(a)

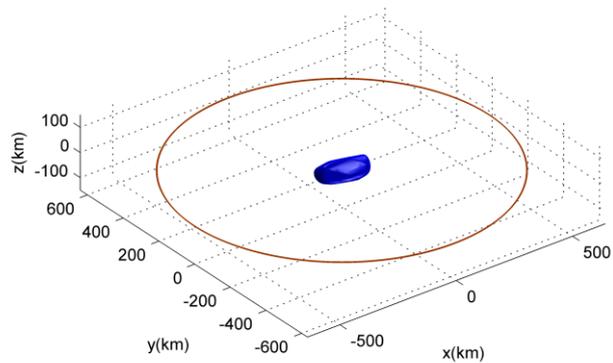

(b)

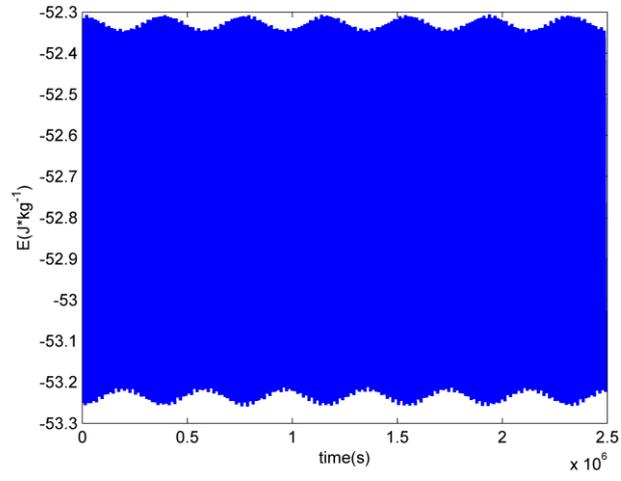

(c)

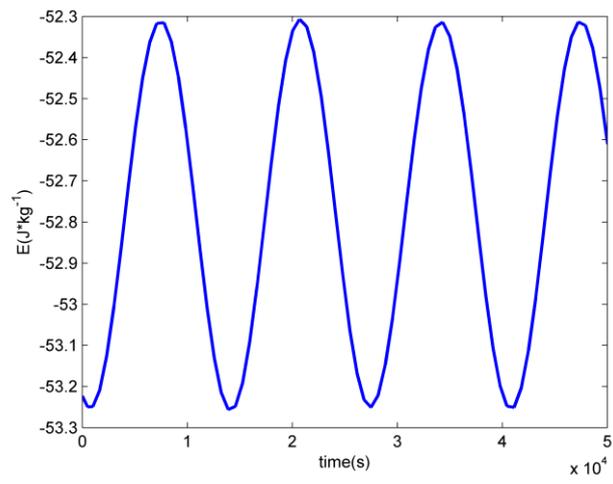

(d)

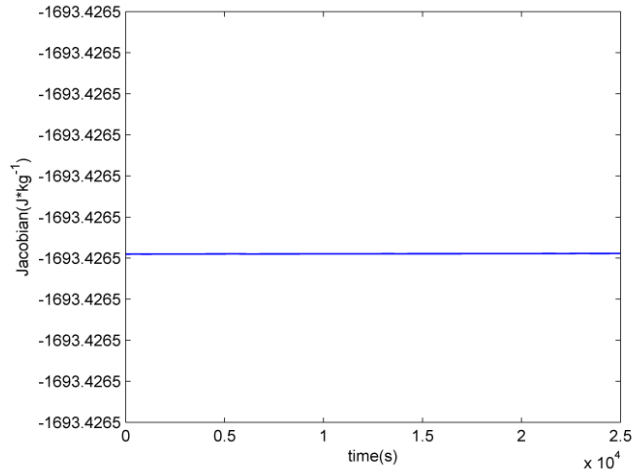

(e)

Figure 8. The calculation of the orbit of the moonlet in the potential of 624 Hektor. (a) The orbit relative to the body-fixed frame and the shape of 624 Hektor; (b) The orbit relative to the inertia system and the shape of 624 Hektor; (c) The mechanical energy of the orbit; (d) Local plot of the mechanical energy of the orbit; (e) Jacobi integral of the orbit.

**4 Conclusions**

The dynamical behaviors in the gravitational potential of the L4-point Jupiter binary Trojan asteroid 624 Hektor have been investigated with the polyhedron model generated by the shape observational data of 2038 faces and 1021 vertexes. Previous literature presents three different density values for 624 Hektor, i.e. 2.43 g·cm$^{-3}$ (Descamps 2014), 1.63 g·cm$^{-3}$ (Carry 2012), and 1.0 g·cm$^{-3}$ (Marchis et al. 2014). We investigated the positions, Jacobian, eigenvalues, topological cases, stability, and the Hessian matrix of the equilibrium points in the gravitational potential of 624 Hektor with different density values. The number, topological cases, and the stability of the equilibrium points with different density values are the same. The positive definite or not and the positive/ negative index of inertia of the Hessian matrix of the equilibrium points remain unchanged when the density varies.

The positions of the equilibrium points vary when the density value of the 624 Hektor varies. If the density value of the 624 Hektor decreases, the distances between the outer equilibrium points (including E1-E4) and the mass center will also decrease. Among these three cases of density values, the outer equilibrium points of 624 Hektor are the farthest relative to the mass center of the body when the density is $\rho=2.43$ g·cm$^{-3}$. The outer equilibrium points move away the mass center of the asteroid as the density increases. The conclusion for the inner equilibrium point E5 is different. The inner equilibrium point E5 moves close to the mass center of the asteroid as the density increases.

Two periodic orbits with similar shapes and different positions are calculated. The results show that there exist unstable periodic orbits near the surface of 624 Hektor. An orbit near the primary's equatorial plane is calculated. The orbit relative to the body-fixed frame has larger amplitude than the orbit relative to the inertia system. The orbit continues to be stable after 28.8375 d.

**Acknowledgements**

This research was supported by the National Natural Science Foundation of China (No. 11772356) and China Postdoctoral Science Foundation- General Program (No. 2017M610875). We would like to thank LetPub (www.letpub.com) for providing linguistic assistance during the preparation of this manuscript.